# Integrating Cognitive Load and Embodied Cognition Theories Through Representations as Multi-Scale Attractors

David C. Gibson
Curtin University, UNESCO Co-Chair
Perth, Western Australia
Australia
davidcgibson50@gmail.com

Mary Elizabeth Azukas
Georgia Institute of Technology
Atlanta, Georgia
United States
Elizabeth.Azukas@gtri.gatech.edu

Meryem Yilmaz Soylu
Georgia Institute of Technology
Atlanta, Georgia
United States
meryem@gatech.edu

Correspondence: davidcgibson50@gmail.com

## Abstract

This article proposes a formal rapprochement between cognitive load theory and embodied cognition by reconceptualizing psychological representations as dynamic multiscale attractors within a temporal-hierarchical prediction architecture. The apparent conflict between the two theories dissolves when viewed through a complex systems lens. Cognitive load theory describes compressed representations operating at medium timescales, while embodied cognition describes fast sensorimotor loops. These two theories describe complementary, timescale-separated processes that operate simultaneously without contradiction. Drawing on dynamical systems theory, hierarchical predictive processing, and a six-node open-systems architecture, the article proposes that learning is best understood as attractor sculpting across coupled temporal layers, from millisecond sensorimotor loops through seconds-to-minutes working memory compression to the slow, years-long reshaping of knowledge structures. Three theoretical reconciliations are developed: time-scale separation, spatially extended hierarchies, and developmental trajectories from novice to expert configurations. From these understandings, five novel, testable predictions are advanced concerning cross-timescale interference, embodied load reduction, metacognition as timescale coupling, feedback topology, and the schema flexibility paradox. For each prediction, converging empirical evidence is reviewed, and formal empirical research designs are proposed. Implications for instructional design, assessment practice, and educational leadership are developed throughout, grounded in the principle that cognitive load and embodied engagement are not competing demands but complementary expressions of a unified temporal-hierarchical cognitive system.

## A Brief Tutorial on Complex Systems and Multiscale Attractors

A complex systems perspective on the integration of cognitive and embodied explanations of learning processes provides a grounding for *rapprochement* between cognitive load (CLT) and embodied cognition (EC) theories of learning processes. Two inseparable and interactive features of complexity serve as mechanisms for integrating the two psychological theories - time and hierarchy. The unification of the two cognitive theories into a temporal-hierarchical constituency is proposed as the primary topographical-topological (physical and semantic) organization for the dynamic and evolving structures that compress experience into memory and knowledge. Supporting evidence for this perspective has been emerging since the late 1990's in learning sciences, computational neurology, and psychology (Bar-Yam, 1997; Baum, 2003; Gell-mann, 1995; Holland, 2019; Rubinov & Sporns, 2010).

Imagine trying to predict a conversation by analyzing a single word or understanding a river by measuring one droplet of water. This is essentially what happens when we apply traditional statistics to living, breathing systems that unfold over time. To understand complex systems and the concept of 'multiscale attractors' that emerges from that perspective requires recognizing three fundamental limitations of conventional statistical approaches. First, statistics typically capture snapshots rather than processes, analyzing one moment in time rather than how a system evolves and transforms. Second, statistics assumes that measurements are independent of one another, when most interesting phenomena involve tangled webs of influence where A affects B, which circles back to affect A (Kelso, 1995). Third, statistics assumes stability, that the variation we measure today will behave similarly tomorrow; but complex systems are characterized by their capacity for change at every level (Kauffman, 1993).

So, what is the alternative? Complex systems science offers a different lens: dynamical systems theory. Rather than asking 'what is the average?' the question is 'where does this system tend to go and how does it behave along the way?' Think of a marble rolling around a bowl. No matter where you place it on the rim, it will eventually settle at the bottom. That bottom point is an *attractor*, a state toward which a dynamic system naturally gravitates (Thelen & Smith, 1994). But real-world systems are far more intricate than marbles in bowls, so attractors come in three varieties: *points* (like the bottom of the bowl), *cycles* (like the sleep-wake cycle), and *chaotic* (like explosions, or somewhat less chaotic, like someone staggering home). In addition, in complex systems, these types of trajectories might be mixed and matched and overlap each other at a variety of levels; in other words, they can be *multiscale*.

Consider, for example, your daily routine. You might have a morning attractor involving coffee, shower, breakfast that plays out over 30 minutes, nested within a weekly attractor of work patterns from Monday to Friday, which itself sits within seasonal attractors like summer vacations and winter holidays. Each operates on its own timescale, yet they influence one another. This is a *multiscale attractor set*, a coherent pattern of behavior that exists simultaneously across different temporal scales, creating stability at multiple levels while remaining flexible enough to adapt (Kelso & Engstrøm, 2006; Van Orden et al., 2003). This is the core idea of the proposed rapprochement between cognitive load theory and embedded cognition that unifies them into one framework.

In biological and social systems, multiscale attractors explain how your heartbeat maintains rhythm from second-to-second while also adjusting to your exertion or your circadian cycle, or how a forest ecosystem maintains stability from day to night while also responding to seasonal changes. These patterns aren't captured by a single measurement or average; they're revealed in how the system moves through time, creating recognizable 'shapes' or 'topologies' of behavior at multiple scales simultaneously (Prigogine & Stengers, 1984). This is the picture we want you to keep in mind as we propose and discuss a reconciliation between cognitive load theory and embedded cognition in which 'representations' in cognitive neuropsychology are understood as multiscale attractors.

Before we can explore how these ideas apply to cognition, learning, and schooling, we need a few more conceptual tools. When a system is tracked over time, its behavior can be plotted in what's called a *state space*; imagine a map where each point represents a possible condition of the system (Strogatz, 2015). A bouncing squash ball, for instance, has a state space defined by its position and velocity in a squash court. As the ball bounces, hits walls, and eventually settles, it traces a path called a *state space trajectory* through this space, which is the route the system takes from one position to another over time. Think of this like tracking a hiker's journey through mountainous terrain. The landscape itself represents all possible positions (the *topology of the state space*), while the actual path the hiker takes is a specific *trajectory*, winding around obstacles, following valleys, and climbing peaks. In complex systems, these trajectories aren't random; they're shaped by the underlying landscape's features, like the walls and net line of the squash court. Complex system landscapes can change over time, imagine the valleys deepening with repeated use, like a path worn into stone, a process called *attractor sculpting* (Schöner, 2008). We are proposing that learning be understood as attractor sculpting in a cognitive state space.

Finally, systems rarely operate at just one speed. A conversation, for example, involves rapid neural firing on the order of milliseconds during listening, word-by-word speech production on the order of seconds, and long-term relationship development with a conversational partner that might take weeks, months, or years. These different timescales don't just coexist; they interact and co-evolve. Fast processes generate patterns that slow processes compress and preserve (see Baum (2003) for a computational cognitive science explanation of knowledge as compression, and what Langer (1942) called 'symbolization'). Conversely, slow processes that form memories and other structures create constraints that shape the unfolding of fast processes, a phenomenon called *temporal stratification* (Newell, 1990; Simon, 1962). Understanding how these temporal layers couple together is essential for grasping multiscale attractors.

To aid in understanding the complex system model of the rapprochement, major cognitive and metacognitive theoretical frameworks can be unified by a six-node network architecture based on open systems and design science theory called the DOT framework (Azukas & Gibson, 2025): **Environment** (external context and task demands), **Inputs** (information entering the learner's system), **Processes** (active cognitive/metacognitive operations), **Structures** (stable and quasi-stable knowledge representations, beliefs, standards), **Output**s (generated products and behaviors), and **Feedback** (information about output quality and environmental response). This architecture provides a common representational network language (with nodes labelled E, I, P, S, O, F, connected by causal relationship linkages) for mapping and analyzing diverse theoretical propositions concerning the dynamic attractor landscape of the rapprochement. The taxonomy of the framework is not only a descriptive tool illustrating the context of the proposed reconciliation, but its pathways (the causal and influential trajectories from node to node) also make visible and mappable the timescale separations and hierarchical couplings that are discussed below.

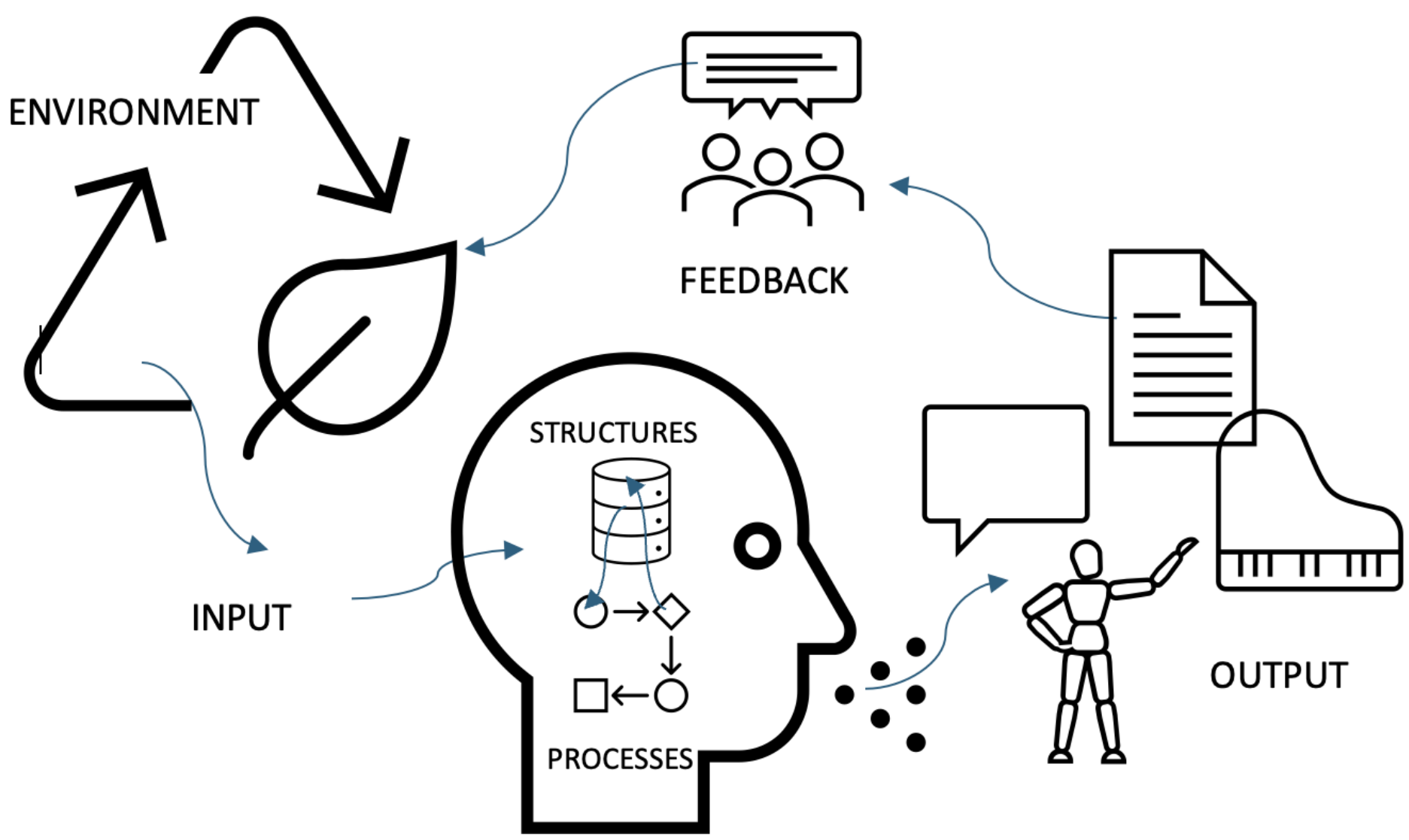

NOTE: The DOT framework's open-system model of cognition has six nodes. Four external (OUTPUT, FEEDBACK, ENVIRONMENT, INPUT) and two internal (PROCESSES and STRUCTURES), and seven primary links or relationships, which form a taxonomy of cognitive processes.

## 1. Temporal Stratification of Cognitive Processes

Human cognition doesn't operate at a single speed; it unfolds across radically different timescales that interact in specific ways. At the fastest layer (milliseconds to seconds), are embodied sensorimotor loops where perception and action couple below conscious awareness (Varela et al., 1991). When you catch a falling glass, for example, your hand is already moving before you've consciously registered the event. This is where cognitive load theory's 'intrinsic load' (Sweller et al., 2019) manifests as *real-time element interactivity*, the immediate demand of coordinating multiple changing relationships. In the DOT framework, the processing node (P) is a nexus of such real-time interactions with the world, conceived not as isolated cognitive computations but as enacted (*re-presented*) patterns emerging from sensorimotor coupling, which is envisioned as the first layer of *embedded cognition*.

At medium timescales (seconds to minutes), storing experiences into working memory and symbolizing experiences into compressed symbols both function as temporal-hierarchical compression mechanisms (Baum, 2003; Langer, 1942). Rather than storing static symbols, working memory and dynamic symbolization actively maintain dynamically adaptive predictions across longer durations, constantly updating to minimize prediction errors (Friston, 2010; Holland, 1998). What cognitive load theory identifies as working memory limitations reflects the system's capacity for sustaining these compressed predictions. The bidirectional relationship between DOT's processing and structures nodes (P⇄S) represents precisely this fast-slow coupling: fast online processing (P) generates (→) slower compressed predictions (S) that then constrain (→) how future processing (P) unfolds.

At the slowest timescales (minutes to years), learning operates as *attractor sculpting*, as defined above. What cognitive load theory calls ‘schemas’ are thus understood as dynamic *hierarchical attractor landscapes* shaped by experience (Thelen & Smith, 1994). The ‘enactive history’ emphasized in embodied cognition becomes formalized as dynamically maintained trajectory-dependent reorganizations of state space; the system literally reshapes its own landscape through repeated patterns of activity. DOT's theory of the structures node (S) conceptualizes short and long-term memories as dynamic potentials and ongoing reorganizational network probabilities rather than static repositories of solid-like ‘representations’ where information sits waiting for retrieval. This reconceptualization underpins the rapprochement and leads to a dynamic view of representation as *re-presenting* past experiences as active, dynamic, temporal hierarchical cognitive compressions that have contextually attached meanings and semantic network relationships, created by co-engaged and excited nodes in the current network, or *connectome topology* (Fornito et al., 2019). The complex systems viewpoint thereby opens new avenues for dynamic description, analysis, testing, and understanding of the mechanisms across the multileveled ‘real world-to-brain-to-mental model’ system.

Each timescale maps onto recognizable moments in educational experience. At the fastest layer, a student learning to read music, for example, navigates sensorimotor loops that control finger placement, breath, and posture, which often operate below the threshold of conscious attention. At medium timescales, a student working through a multi-step proof holds partial results in working memory while compressing them toward a solution. At the slowest timescale, a student who has spent three years studying history gradually develops the disciplinary attractor landscape containing interpretive schemas, recurring questions, and habits of evidence that constitute genuine historical thinking. Recognizing that all three processes are simultaneously active in any learning episode reframes the teacher's task, not to manage a single cognitive resource, but to support learners across all three temporal layers at once.

## 2. Hierarchical Predictive Processing as a Bridging Framework

Rather than viewing mental representations as *static symbols,* as in cognitive load theory’s schemas, or *eliminable constructs* in embodied cognition as advocated by Brooks, (1991), the complexity framework sees representations as *temporally-scaled attractor dynamics* within *hierarchical prediction architectures*. This bridges the two psychological frameworks (cognitive load theory and embodied cognition) through three key physically constrained moves: 1) temporal stratification, which leads to the concept of time scale separation, 2) hierarchical predictive processing (Fornito et al., 2019), which distributes cognitive processing over both time and spatial separations, and 3) developmental or ontological mechanisms in which cognitive regimes are dynamically stable in the short term but also evolve over longer time periods.

How does this work? According to Hawkins, (2021) a cognitive hierarchy level predicts the level immediately below itself and is constrained by that lower level's prediction errors. This basic mechanism of neuroscience resolves the apparent contradiction between cognitive load theory and embodied cognition approaches. Cognitive load is reinterpreted as precision-weighted prediction error at multiple hierarchical levels. For example, intrinsic load is formed by irreducible prediction error at task-relevant timescales. Extraneous load is created by misdirected predictions from poorly designed or functioning interfaces. Germane load is understood as productive error (see Jacobson et al., (2019) and Kapur (2015) for a discussion of productuve failure in learning) that reshapes higher-level attractor landscapes in the process of schema construction, or attractor shaping.

In terms of embodied cognition, prediction isn't just 'in the head,' it is spatially distributed across brain-body-environment loops (Lakoff & Johnson, 2003; Varela et al., 2017). Lower levels predict sensorimotor expectations and consequences. Higher levels predict lower-level states, which are abstracted from the sensorimotor layers and thus constitute a hierarchical component of the dynamic re-presentation that is open to adaptation over time by generating exception rules (see Holland, (2019) for information about the *quasihomomorphism mechanism* of adaptation). In embodied cognition, the body becomes part of the predictive architecture; it is not an external input to some other cognitive mechanism. This implies, as Brooks (1991) pointed out, that *cognition is distributed* and does not require a 'central processing unit.' In the complexity-based rapprochement, cognition is something accomplished by a 'system' (which includes externalized processes and representations) rather than solely by a centralized, internalized brain function.

The DOT framework's six-node network of the proposed theoretical reconciliation models the complexity view of temporal-hierarchical and spatially distributed prediction. The internal nodes (P and S) mappings include the P-node: Fast online prediction generation at the sensorimotor level (Brooks, 1991); S-node: Slower attractor landscapes that generate top-down predictions (Hawkins & Blakeslee, 2004); P→S: Prediction error signals that sculpt attractors (Holland, 1998); S→P: Top-down control and predictions that constrain processing (Thagard, 2005); and P⇄S: a crucial bidirectional coupling across timescales (Xiong, 2022). Complementing these trajectories, the external nodes mappings (E, O, F, I) include Environmental regularities that scaffold hierarchical prediction (Gigerenzer & Todd, 1999a; Kahneman, 2011), Outputs that provide externalized representations with physical and psychological distance (Johnson et al., 2006; Thagard, 2010), Feedback that separates consequences from actions and learning (Hattie & Temperley, 2007; Ifenthaler, 2011), and Inputs that can disequilibrate one's dynamic homeostasis (Piaget, 1985).

The core ideas of cognitive load theory (intrinsic, extraneous, and germane loads), understood through the DOT framework's temporal-hierarchical architecture, are viewed as types of prediction errors derived from the dynamics of the learning process. Intrinsic load is the irreducible prediction error generated when task elements operate across timescales that cannot yet be hierarchically compressed or symbolized by the learner. Extraneous load is the prediction error introduced when environmental design misaligns with the learner's current coupling structure, forcing processing at the wrong timescale. Lastly, germane load is the productive error that arises precisely at the boundary between existing attractor landscapes and new experiences, driving the P→S sculpting that constitutes learning itself.

These reinterpretations of load types have direct instructional implications. Intrinsic load sets a floor below which simplification cannot go without distorting the discipline. A well-designed problem retains this load because grappling with it is required for learning. Extraneous load, by contrast, signals a design failure, for example, a poorly sequenced lesson, a cluttered diagram, or misaligned assessment instructions that generate prediction errors unrelated to the learning target. Germane load is what a teacher cultivates when they introduce a carefully chosen counterexample, a disequilibrating case, or a problem designed to fail before it succeeds. The pedagogical art lies in distinguishing all three in real time.

## 3. Theoretical Rapprochements Enabled

As a result of the above considerations, supported by a wide range of research-based scholarship, the ensuing integration of theories treats psychological representations as dynamic multiscale attractors in a neuropsychological landscape, which in turn leads to three mechanisms of reconciliation: time-scale separation, spatially extended hierarchies, and developmental trajectories. The following sections briefly define and describe the three resolutions and point to implications for helping students to learn and teachers to teach, which also sets a challenging agenda for educational leadership founded in the learning sciences.

### 3.1 Time-Scale Separation

The apparent conflicts of cognitive load theory versus embodied cognition dissolve when we recognize that cognitive load theory describes *compressed representations at medium timescales* while embodied cognition's foundations begin with *fast sensorimotor loops*. Both are necessary and are timescale separated, so there is no conflict when simultaneously activated. Working memory limits in cognitive load theory emerge from the bandwidth constraints of compressing fast dynamics into slower predictive models, while embodied scaffolding offloads prediction to environmental regularities, reducing compression requirements. Gesture and spatial arrangements, for example, lower cognitive load due to time-scale separation (Goldin-Meadow et al., 2007). In this interpretation, cognitive load reflects the resource cost of sustaining temporally extended predictive compressions under bandwidth constraints. When lower-level sensorimotor prediction errors cannot be resolved through embodied-environmental coupling, existing predictive structures fail to assimilate incoming information. At this threshold, accommodation processes are required to reorganize the attractor landscape (Gibson & Ifenthaler, 2024). This situation translates into measurable constructs, including increased reaction time, error rates, physiological stress, a breakdown of smooth action, and disfluency signals.

From the perspective of timescale separation, the core tenets of cognitive load theory are thus understood as arising from the system's dynamics. Intrinsic load reflects the compression cost of coordinating elements that simultaneously span multiple temporal layers. Extraneous load reflects the additional compression cost imposed when embodied scaffolding is absent, poorly timed, or misaligned with the learner's current sensorimotor capabilities. Lastly, germane load reflects the productive compression work occurring at the assimilation-accommodation boundary, where fast sensorimotor dynamics are being transformed into slower, more stable attractor structures.

These theoretical distinctions carry immediate practical weight for instructional design and assessment across educational levels. A science teacher who asks students to physically manipulate molecular model kits before introducing symbolic chemical notation is not simply adding a hands-on activity for engagement; she is strategically offloading prediction to environmental regularities (the physical affordances of the kit) before requiring learners to compress those dynamics into the symbolic representations that chemical notation demands. Physical gestures precede the creation of a symbol because the sensorimotor loop must be established before the compression cost is justified. Similarly, a university lecturer who requires students to sketch a diagram of a system before explaining it in writing is sequencing embodied scaffolding ahead of representational load, not as a pedagogical preference but as a temporal, developmental necessity.

Assessment practices aligned with this principle distinguish between tasks that probe embodied fluency (Can the student navigate the physical or spatial problem with ease?) and tasks that probe representational compression (Can the student explain, predict, or transfer using abstracted models?). Treating these as equivalent or interchangeable, as when a standardized test bypasses embodied engagement entirely, systematically disadvantages learners who are at the assimilation-accommodation decision point Piaget describes. A standardized test is structurally incapable of measuring the knowledge and skills of those for whom environmental scaffolding has not yet reduced compression requirements enough to make symbolic performance reliable.

For educational leaders, this framework reframes the recurring debate about manipulatives, physical learning environments, and the spatial organization of classrooms not as questions of preference or resource allocation, but as questions about whether the institutional environment provides adequate embodied scaffolding to reduce the compression burden on working memory. A school that strips gesture, movement, and spatial arrangement from its pedagogy in pursuit of efficiency is, in the terms of this framework, artificially inflating the cognitive load for all learners and removing the very mechanism by which assimilation failure is most naturally resolved. A reflective educational leader might ask: *Where in our current unit designs or assessment sequences are we implicitly requiring representational compression before adequate embodied scaffolding has been established, and how would we know?*

## 3.2 Extended Hierarchies

The transformation of experience into knowledge (sensorimotor processes becoming dynamic embodied structures that can be re-membered and applied to solve problems) is internal to the learner but not isolated from the environment. The transformation is the brain's contribution to a broader network of prediction and control that includes distributed resources and overlapping semantic layers. The externalized nodes of the DOT model that provide input, outputs, feedback information, and influence the broader environment, are not merely inputs and outputs but are *extended prediction loop components* through the body *embedded within* its environment. This line of thinking leads to seeing cross-cluster connections as hierarchical couplings where internal predictions engage environmental regularities, as indicated in the simple heuristics that make us smart (Gerd Gigerenzer & Todd, 1999) and (Reimer et al., 2006).

Consider, for example, improvising musicians in a jazz cafe, where each note a soloist plays immediately influences the choice of the next note via a fast, internal dialogue of self-monitoring (what DOT represents as O→I, output directly linked to input). Meanwhile, a responding musician's reply might support or interfere with the soloist's choices, creating a medium-speed feedback loop of environmental influence (F→I, feedback signal directly linked to input). On a slower timescale, the combined sounds of both musicians fill the cafe, layering over the background conversations of customers and shaping the room's ambience, a longer-term influence that sculpts the attractor landscape of audience appreciation and the band's task environment (O→E, output directly linked to the larger environment). The DOT model's taxonomic shortcuts (O→I or O→F instead of O→F→E→I) represent additional routes through the extended hierarchy, cycling cognitive products, like the jazz expressions, back to input. The shortcuts are theoretically meaningful, especially to experts, as different temporal-hierarchical couplings: O→I operates at the speed of immediate perception-action; F→I at the speed of interactive exchange; and O→E at the speed of environmental transformation. Likewise, cognitive embodiment and cognitive load operate at different speeds and levels of an extended hierarchical cognitive system. While we refer to the alternative trajectories in the external nodes of the DOT model as 'shortcuts' (because they dispense with the longer causal path from an output becoming feedback for the environment and eventually becoming new input to the learner), we could also think of them as feedback elaborations. Experts make use of many more of these shortcuts for self, peer, and community improvement information, compared to novices.

The jazz cafe illustration makes vivid something that instructional designers and educational leaders often handle intuitively but rarely articulate with precision: learning environments are not simply containers for cognition but are active, dynamic influences in hierarchical prediction loops operating simultaneously across different timescales. A professional learning community, for example, is not merely a structure for sharing practices; it is an extended prediction network in which a teacher's classroom actions immediately reshape their own next instructional move, their interactions with peers influences their thinking, and the norms, physical arrangements, scheduling patterns, and leadership tone of the school shapes the cultural landscape within which the faster activities occur.

Contemporary instructional technologies provide concrete instantiations of extended hierarchical coupling. Immersive virtual environments, for example, sustain perceptual access to relational structures across time, reducing reliance on limited working-memory capacity. Reviews indicate that immersive environments are particularly beneficial when learners must coordinate complex spatial information (Jensen & Konradsen, 2018; Radianti et al., 2020), and when well designed, VR reduces extraneous cognitive load by integrating interacting elements and minimizing split-attention demands (Makransky & Petersen, 2021; Mutlu-Bayraktar et al., 2019). Within the multiscale attractor framework, immersive systems function as external scaffolds that stabilize prediction structures by externalizing representational demands. In doing so, part of the coordination burden shifts from internally sustained working memory to persistent perceptual supports embedded in the environment.

Educational and adaptive systems extend this principle by structuring biologically secondary information in ways that reduce element interactivity and unnecessary coordination demands (Mayer, 2021; Sweller, 2020; Sweller et al., 2019). Instructional procedures that integrate interacting elements, provide explicit guidance, and scaffold sequencing decrease the number of elements that must be processed simultaneously, improving learning efficiency (Deva et al., 2025). Adaptive systems further externalize metacognitive monitoring and control by dynamically adjusting feedback based on learners' cognitive and affective states (Azevedo et al., 2019). Regulation thus becomes distributed across learner–system interactions rather than sustained internally, exemplifying extended hierarchical coupling in which external feedback loops contribute to the stabilization of representational dynamics.

Embodied interaction reflects the same architectural principle. Bodily action and environmental engagement operate as mechanisms of cognitive offloading, supporting representational integration that would otherwise rely on sustained working-memory coordination (Castro-Alonso et al., 2024; Risko & Gilbert, 2016). Meta-analytic evidence indicates that embodied interaction reduces cognitive load while improving performance (Lyu & Deng, 2024). Within the present framework, embodiment stabilizes prediction states through continuous perception–action coupling, reducing internal compression demands while preserving fast sensorimotor dynamics.

Lesson and unit design become richer when deliberately planned across multiple timescales. In-the-moment formative checks and physical or gestural prompts might serve the fast loop. The medium loop might be supported by structured peer dialogue, iterative drafting, or spaced retrieval cycles, and the slow loop may be served by the cumulative semantic environment a unit builds via vocabulary on the walls, recurring problem types, and disciplinary habits that accrue and are honed across weeks.

Assessment aligned with this model would sample across timescales rather than collapsing everything into a single summative measure. What is needed are observations of real-time decision-making under constraints, examinations of how learners (and teachers) use feedback across a task sequence, and evaluations of how their behaviors shift: what resources people reach for automatically, what framings they spontaneously apply, and how these actions change over time.

An educational leader who intervenes to change the macro environment based only on 'outcomes' (for example, by restructuring timetables, redesigning professional learning formats, revising curriculum frameworks) without also attending to the faster loops will find that environmental transformation is slow to penetrate the landscape of classroom practice. Conversely, a teacher who works only at the level of immediate feedback during instruction, for example, 'teaching to the test,' risks missing how medium-speed institutional feedback loops are reinforcing or undermining the very moves they are trying to develop. A reflective educational leader might ask: *In our current professional learning design or unit planning, which timescale receives the most deliberate attention, and which is being left to chance?*

### 3.3 Developmental Trajectories

The DOT framework's taxonomic approach to trajectories computationally defines configurations of temporal-hierarchical prediction capabilities that people develop as novices, apprentices, and eventually experts (Gibson & Azukas, 2025). Novice scenarios rely heavily on fast sensorimotor loops (P-dominant), with minimal compression into slow attractors (weak or rigid S). Mid-development scenarios are characterized by emerging P⇄S bidirectionality as learners develop compressed predictions (schemas, attractors, symbols, and meaning) from experience, and are most likely to have their fragility misread as inconsistency or unreliability rather than as a structural developmental condition. Expert scenarios involve more sophisticated multi-timescale integration where dynamic attractor landscapes (flexible S) efficiently predict without overwhelming slow conscious processing. This characterization is consistent with cognitive load theory's reduced load through expertise. These newly detailed development trajectories further explain why experts exhibit both *reduced cognitive load* due to efficient temporal compression of well-sculpted attractors and *enhanced embodied engagement* via fast loops operating more fluidly without conscious mediation. Experts, in this perspective, have richer continuous feedback utilization because they make active simultaneous use of multiple timescales and hierarchical couplings; for example, they can keep a bigger picture in mind while attending to details compared to a novice, who may be flooded by details.

The developmental architecture, which resolves some of the conflicts between cognitive load and embedded cognition theories, reframes both the diagnosis of struggle and the design of progression in ways that cut across all educational levels. A novice learner who appears overwhelmed, distracted, or unable to reflect on their own performance is not simply lacking effort or metacognitive awareness. They are operating in a process-dominant configuration in which fast sensorimotor loops are consuming the available bandwidth, leaving little or no capacity for the slower, compressed predictions that self-monitoring and strategic adjustment require. Instruction designed without this understanding will frequently misread sensorimotor saturation as motivational deficit. An apprentice learner, by contrast, is precisely at the most demanding developmental juncture. Bidirectionality is emerging, which means they are simultaneously managing fast perceptual demands while also beginning to form slower compressed representations, a double load that explains the characteristic fragility of intermediate performance. These students are competent under familiar conditions, but their performance is rapidly degraded under novel ones. Expert performance, on this account, is not the absence of sensorimotor engagement but its liberation. Fast loops operate fluidly beneath conscious mediation precisely because slow attractor structures are robust enough to carry the predictive weight, freeing attention for the multi-timescale integration that allows an expert to hold the big picture while simultaneously tracking fine-grained detail.

Mapped onto the DOT developmental configurations, the core tenets of cognitive load theory can be explained in new terms. Intrinsic load defines the complexity ceiling of a given stage, whether novice, apprentice, or expert. For example, each stage of development has an intrinsic load limit beyond which the learner cannot further compress (symbolize in relation to previous knowledge) without first developing richer coupling between automated processes and newly accommodated knowledge. Extraneous load at any stage imposes cross-timescale and hierarchical organization interference beyond what the learner's current configuration can handle, which may also incidentally include interference from actual incongruities and incompatibilities inherent in the real world. Finally, dealing successfully with germane load expands the learner's attractor landscape without overwhelming the temporal bandwidth available at that developmental juncture.

For lesson and unit design, understanding the developmental trajectories at the intersection of cognitive load and embedded cognition suggests that appropriate sequences are not only 'easy-to-hard' or 'concrete-to-abstract' but also 'process-scaffolded to bidirectionality-supported to multi-timescale integration.' Students as well as teachers need explicit transitional support at the apprentice stage, where the compression cost is highest, and the risk of over-generalized, revision-resistant schemas is greatest.

Assessment and feedback practices calibrated to these developmental trajectories look different at each stage. For novices, feedback is needed that reduces sensorimotor load and anchors fast loops in environmental regularities. For apprentices, feedback is needed that explicitly bridges embodied experience and emerging compressed representations, helping learners notice when their schemas are failing to predict. For experts, feedback should interrogate the person's flexibility under genuinely novel demands; the issue is not whether they can perform, but whether their performance degrades gracefully or catastrophically under pressure.

Educational leaders designing professional learning, curriculum progressions, or induction programs for new teachers would do well to map their cohorts against these configurations rather than treating developmental diversity as variance to be normalized. The novice teacher flooded by classroom detail, the mid-career teacher whose schemas have become rigid and resistant to new evidence, and the expert teacher who navigates complexity with apparent ease are not points on a single continuum of competence but qualitatively different configurations of temporal-hierarchical coupling. Each professional stage requires a different kind of environmental scaffolding, feedback, and attractor-expanding challenge to develop further. A reflective educational leader might ask: *When I observe a learner or a teacher who is struggling, what evidence would help me distinguish between **sensorimotor saturation**, **schema fragility** at the apprentice transition, or **mental rigidity** in an experienced performer, and does my current feedback practice make that distinction?*

## 4. Theoretical Predictions

The conceptualization of representation as multiscale attractors generates five novel testable predictions, which the remaining sections will examine.

> *Cross-timescale interference*: Cognitive load should increase when task demands require simultaneous attention to multiple temporal scales that cannot be hierarchically organized.
>
> *Embodied load reduction*: Physical scaffolding (gesture, spatial arrangement, tool use) should specifically reduce load at medium timescales (working memory) by offloading temporal compression requirements but may not affect fast sensorimotor processing.
>
> *Metacognitive development as time-scale coupling*: Perception sending signals for monitoring by higher levels (via pattern finding and making) and higher levels, in turn, controlling perception (via attention) should show increasing temporal sophistication with expertise. That is, experts can monitor and control across wider temporal spans.
>
> *Feedback topology and temporal coupling*: Shortcuts in information processing (outputs direct to inputs without feedback, or environment absorbing outputs as feedback) should map to specific temporal coupling patterns. For example, a person utilizing more shortcuts should exhibit richer multi-timescale integration. Such a person should have an increased ability to reshape environmental attractors at slow and longer timescales (e.g. should be able to manage more complex innovation processes).

*Schema flexibility paradox*: Cognitive load theory's 'expertise reversal effect' occurs when highly compressed attractors (expert schemas) cannot flexibly respond to novelty, which is explained by over-optimization for specific temporal predictions.

These five predictions are not merely laboratory hypotheses; each one carries a distinctive implication for how educators design learning environments, interpret student behavior, and lead institutional change. The cross-timescale interference prediction suggests that a student who appears cognitively overwhelmed may not simply be encountering difficult content, they may be caught between simultaneous demands operating at incompatible temporal scales, such as tracking fast-changing worked examples on a screen while simultaneously trying to maintain a slower, schema-level understanding of the underlying principle. Recognizing this as a structural problem of temporal misalignment, rather than a motivational or ability deficit, changes the instructional response. The remedy is temporal scaffolding, not simplification.

Embodied load reduction reframes the pedagogical value of gesture, spatial layout, and physical materials not as motivational enhancements but as load-management tools that specifically target the working-memory compression bottleneck, leaving fast sensorimotor loops intact and available. This gives principled justification for embodied practices that are often defended on engagement grounds alone.

Metacognitive development as a time-scale issue implies that a novice student's (or teacher's) poor self-monitoring is not a character trait or a skills deficit but a structural consequence of operating within a narrow temporal window. Instruction (or professional development) designed to gradually widen that window, by helping people connect immediate performance signals to longer-range patterns, develops the temporal architecture of metacognition itself.

The variety of paths of feedback carries perhaps the most direct implications for educational leadership. A school or university that invests in rich, multi-layered feedback infrastructures (formative assessment systems, reflective portfolios, mentoring relationships, peer review cycles, and data dashboards that operate across different timescales) is improving feedback quality by expanding the temporal depth and reflective complexity from which learners and teachers can reshape their agency for self-assessment and improvement.

Finally, the schema flexibility paradox reframes one of the most persistent puzzles in professional learning: why are experienced teachers sometimes the hardest to develop, and the most capable students the hardest to teach? An expert teacher whose schemas are highly compressed and efficient under familiar conditions may be exhibiting attractor rigidity (rigid mental models in this case), not resistance or unwillingness, when confronted with genuinely novel instructional demands. This reframing shifts the leadership response from persuasion toward designing experiences that introduce productive temporal mismatch. The novelty must be structured carefully enough to generate schema-expanding prediction error without overwhelming the expert's (or gifted student's) capacity to reorganize.

In the next sections, we take each of these predictions and provide research-based support to aid in developing empirical tests of the predictions. More detailed outlines of proposed empirical research for subsections 5.1 through 5.6 are provided in Appendix 1.

## 5. Cross-timescale interference: Research support

*Cognitive load should increase when task demands require simultaneous attention to multiple temporal scales that cannot be hierarchically organized.*

### 5.1 Cognitive Load, Element Interactivity, and Temporal Coordination

Cognitive load theory (CLT) establishes that cognitive load increases as element interactivity increases, particularly when elements must be processed simultaneously rather than sequentially or hierarchically (Sweller, 2010; Sweller et al., 2019). Although CLT does not explicitly frame element interactivity in temporal-scale terms, the requirement to coordinate processes operating over distinct temporal spans effectively multiplies element interactivity. Empirical evidence shows that tasks requiring simultaneous processing of fast and slow informational streams produce disproportionate performance decrements relative to tasks where processing can be temporally staged (Ayres, 2006; Leppink et al., 2013). These findings support the idea that *temporal misalignment functions as a hidden source of intrinsic load*. The evidence implies that when different temporal scales cannot be nested or hierarchically organized, working memory must actively coordinate across incompatible update rates, increasing load.

### 5.2 Dual-Task and Psychological Refractory Period (PRP) Evidence

Dual-task and PRP paradigms demonstrate robust interference when two tasks demand concurrent central processing, even when tasks differ in modality (Pashler, 1994; Tombu & Jolicoeur, 2003). Importantly, interference is strongest when tasks require simultaneous central coordination, rather than when one task can be buffered or hierarchically postponed. More recent computational accounts interpret PRP effects as *failures of temporal coordination across processing stages*, not simply resource depletion (Zylberberg et al., 2010). This aligns with the proposed mechanism: interference arises when *processes with incompatible timing constraints cannot be serially or hierarchically arranged*. The evidence here implies that cross-timescale interference is a special case of central bottleneck overload driven by incompatible temporal demands. Experts might solve this problem better than novices due to the more flexible and wider scope of their dynamic attractors.

### 5.3 Hierarchical Temporal Processing in the Brain

Neuroscience provides strong evidence that cognition is organized hierarchically across temporal scales**.** For example, sensory regions integrate information over milliseconds, association areas integrate over seconds, and higher-order regions integrate over tens of seconds or longer (Hasson et al., 2008; Lerner et al., 2011). When tasks align with this hierarchy, processing is efficient. However, when tasks require simultaneous access to multiple temporal receptive windows without clear nesting (tracking moment-to-moment signals while maintaining long-range predictive structure), neural efficiency declines and control regions (frontoparietal networks) show increased activation (Badre & D'Esposito, 2009). This evidence implies that for attentional monitoring, the brain is attracted to *temporal scale separation*; violating separation increases control demands, an operational neural correlate of increased cognitive load. A corollary is that for control, the brain is dependent on *temporal scale integration;* violating integration increases attention, assimilation, and accommodation demands.

## 5.4 Attention to Competing Temporal Features

Experimental work on attention to time-varying features shows that simultaneously attending to multiple temporal structures (e.g., rhythm and duration, fast modulation and slow trends) increases behavioral error and neural control demands (Coull et al., 2013). Crucially, when temporal features can be hierarchically related (when a slower rhythm organizes faster events), interference is reduced. When timing features are orthogonal or competing, interference increases. Therefore, the evidence suggests that *temporal hierarchy is not optional*; it is a load-reducing organizational constraint.

## 5.5 Physiological Load Indicators

Pupillometry and neuroimaging studies consistently show that load increases non-linearly when tasks require maintaining and coordinating multiple concurrent temporal representations (van der Wel & van Steenbergen, 2018; Unsworth & Robison, 2015). These increases exceed what would be predicted by task difficulty alone, suggesting that coordination across timescales introduces an additional, independent source of cognitive cost.

## 5.6 Cross-Timescale Claim Limitations

Expertise research shows that experts can often manage multiple temporal scales with reduced load by compressing slow dynamics into higher-order chunks or schemas (Ericsson & Kintsch, 1995; Gobet et al., 2001). This does not negate the cross-timescale claim but qualifies it with two potentially testable conditions: cross-timescale demands increase load when hierarchical organization is unavailable, and expertise reduces load precisely by restoring temporal hierarchy. Tests of these hypotheses are outlined in Appendix 1.

While the converging evidence is strong, it is important to note that few studies explicitly manipulate *temporal-scale incompatibility* as an independent variable, so most evidence is inferential, drawn from dual-task, hierarchy, and attention research. Thus, the cross-timescale claim is strongly supported theoretically and indirectly but remains under-tested as an explicit experimental construct. Extant research supports a trio of ideas: 1) cognitive systems rely on hierarchical temporal organization, 2) simultaneous demands across incompatible temporal scales increase control demands, and 3) increased control demands manifest as increased cognitive load. Direct experimental paradigms explicitly manipulating hierarchical vs. non-hierarchical temporal coupling are needed, as is the formal integration of temporal-scale separation into cognitive science metrics. Proposed empirical tests for subsection 5 are provided in Appendix 1.

## 6. Embodied load reduction: Research support

*Physical scaffolding (gesture, spatial arrangement, tool use) should specifically reduce load at medium timescales (working memory) by offloading temporal compression requirements but may not affect fast sensorimotor processing.*

### 6.1 Gesture as a Working-Memory Load Reduction Mechanism

A robust literature demonstrates that gesture reduces working memory load, particularly during problem solving, explanation, and learning. Goldin-Meadow and colleagues, for example, have repeatedly shown that producing gestures while explaining a problem reduces concurrent working-memory demands, as measured by improved performance on secondary memory tasks (Goldin-Meadow et al., 2001; Goldin-Meadow et al., 2007). Importantly, gesture does not speed basic perceptual or motor processes, and its benefits emerge when learners must maintain, integrate, or transform information over seconds. Cook et al. (2012) further demonstrated that gesture facilitates learning by externalizing relational structure, allowing learners to offload internal maintenance demands. This evidence suggests that *gesture functions as an external representational medium that reduces the need for temporal compression* in working memory, consistent with our rapprochement claim.

### 6.2 Spatial Arrangement and External Representations

Research on external representations shows that spatial layouts reduce cognitive load by replacing internal sequential processing with perceptual structure. Zhang and Norman (1994) demonstrated that well-designed external representations change the *nature* of cognitive processing by shifting work from internal memory to the environment. Similarly, Kirsh (2010) showed that spatial reorganization of task elements ("epistemic actions") reduces memory-and-control demands without altering low-level sensorimotor execution. In instructional contexts, spatially integrated diagrams reduce extraneous load and improve learning outcomes specifically at the working-memory level, as shown in split-attention and modality-effect research (Chandler & Sweller, 1991; Sweller et al., 2019). This evidence suggests that spatial scaffolding reduces load by stabilizing information over time, allowing perception to replace memory, precisely an offloading of medium-timescale temporal compression.

### 6.3 Tool Use and Cognitive Offloading

Tool use has long been shown to function as a cognitive extension, particularly for memory and planning. Risko and Gilbert (2016) for example, review evidence that humans systematically offload cognitive work onto tools (notes, diagrams, calculators) to reduce internal memory demands. Neurocognitive studies of tool use show that tools become integrated into task-relevant representations, reducing working-memory load while leaving sensorimotor loops intact (Baber et al., 2013). Crucially, these benefits are task-level and representational, not reflexive or perceptual. This evidence suggests that tool use offloads medium-timescale representational maintenance, not fast sensorimotor processing.

## 6.4 Embodied Cognition and Load Reduction in Learning

Meta-analyses and experimental studies in embodied learning show that embodied actions improve learning outcomes primarily when tasks require integration over time. Skulmowski and Rey (2018), for example, found that embodiment effects are strongest when actions map onto conceptual or relational structure, not when tasks are purely perceptual. Likewise, Johnson-Glenberg et al. (2014) showed that embodiment improves retention and transfer by anchoring abstract representations in stable sensorimotor patterns. These research-based results imply that *embodiment aids learning by stabilizing representations across time*, thereby reducing working-memory load.

## 6.5 Timescale Specificity Evidence and Limitations

Fast sensorimotor processes (tens to hundreds of milliseconds) are largely automatic and operate below conscious control. Evidence from motor control and perception research indicates that gesture and tool use do not reduce the computational cost of these fast loops; instead, they recruit them (Wolpert & Ghahramani, 2000). Embodied supports often increase sensorimotor activity while decreasing subjective and objective cognitive load (via lower dual-task interference), demonstrating a dissociation between fast and medium timescales (Goldin-Meadow et al., 2001). These sources indicate that embodied scaffolds *redistribute effort across timescales* rather than uniformly reducing processing demands.

Embodiment does not reliably benefit tasks dominated by simple perceptual discrimination, automatized motor execution, or tasks that are already well-chunked at the sensorimotor level (Skulmowski & Rey, 2018; Wilson & Golonka, 2013). This limitation supports the claim that embodied scaffolding does not primarily target fast sensorimotor processing. Instead, embodied scaffolds reduce extraneous and intrinsic load by restructuring information so that fewer interacting elements must be maintained simultaneously (Sweller et al., 2019). External artifacts and actions function as *temporal stabilizers*, allowing cognition to unfold across agent–environment systems rather than within working memory alone (Hutchins, 1995). Thus, it appears that *embodied scaffolds reduce uncertainty at intermediate predictive horizons* while leaving low-level prediction loops intact

In summary, converging empirical evidence supports the notions that gesture, spatial layout, and tool use reliably reduce working-memory load, that the resulting benefits arise at seconds-to-minutes timescales, and that fast sensorimotor processing is not reduced and may be increased. Direct measurement of 'temporal compression' (creating a symbol at a higher level based on a lower level's patterns) as a mediating variable needs further investigation, as does neural dissociation of fast vs. medium timescale load under various embodiments. The 'multiscale attractor' formulation advanced here seems to integrate results that are typically reported without an explicit integrated timescale and spatial theory.

These timescale-specific findings carry a practical design principle: embodied supports work best when they are positioned to reduce working-memory compression demands, not when they are used to automate already-fluent skills. A mathematics teacher who introduces manipulatives to students who have already internalized a procedure is adding sensorimotor activity without reducing load, because the compression work is already done. The more productive application is earlier in the learning sequence when students must simultaneously track multiple interacting elements, and the compression cost is higher. This implies that lesson design should treat embodied scaffolding not as a general-purpose engagement strategy but as a targeted intervention, deployed at precisely the points where medium-timescale compression requirements would otherwise overwhelm working memory.

## 7. Metacognition as time-scale coupling: Research support

*Perception processes sending signals for monitoring by higher levels (via pattern finding and making) and higher levels, in turn, controlling perception (via attention) should show increasing temporal sophistication with expertise. That is, experts can monitor and control across wider temporal spans.*

### 7.1 Monitoring Across Time

A foundational result in metacognition research is that monitoring accuracy increases with expertise and development, particularly for judgments requiring integration over time rather than momentary confidence. Koriat (1997) demonstrated that effective metacognitive monitoring depends on *cue utilization*, with experts relying more on diagnostic, temporally extended cues (coherence, error patterns) rather than immediate perceptual fluency. Dunlosky and Metcalfe (2009) further showed that novices' monitoring is dominated by short-term cues, while experts can assess learning and performance over longer intervals. In learning contexts, judgments of learning become more accurate when learners delay monitoring, forcing reliance on longer-timescale memory dynamics rather than transient working-memory traces (Rhodes & Tauber, 2011). The evidence suggests that *monitoring of perception from memory shifts from fast, local signals to slow, temporally integrated signals as expertise develops*.

### 7.2 Control Across Time

Metacognitive control, that is, decisions about allocation of effort, strategy selection, and regulation, also shows increasing temporal horizon with expertise. Research in self-regulated learning, for example, demonstrates that novices engage in reactive, short-term control (re-reading immediately after failure), whereas experts engage in prospective and strategic regulation, planning across minutes, hours, or longer learning episodes (Winne & Hadwin, 2008; Zimmerman, 2002). Efklides (2011) distinguishes between object-level cognition via fast processing and meta-level regulation operating at slower, longer timescales. Experts exhibit more effective coupling between levels, enabling sustained regulation across extended tasks. These findings indicate that *top-down control of perception via remembered or dynamically reconstructed structures increasingly operates over longer temporal spans, constraining fast processing with slow, goal-oriented structure*.

## 7.3 Expertise and Temporal Depth of Cognitive Representations

Expertise research shows that experts possess deeply structured representations that allow both monitoring and control over extended sequences of actions. Ericsson and Kintsch's (1995) theory of long-term working memory, for example, demonstrates that experts can access and regulate information across extended temporal spans without overwhelming working memory. This capability enables monitoring performance trends rather than isolated errors and controlling strategy based on long-range task structure. In domains such as chess and medicine, experts anticipate future states and regulate present actions accordingly, a hallmark of slow-timescale control influencing fast processing (Chi et al., 1988).

## 7.4 Neural and Developmental Evidence for Increasing Temporal Integration

Neuroscientific evidence also strongly supports dynamic hierarchical temporal organization in cognitive control. The prefrontal cortex is organized along a front-to-back axis, with frontal regions supporting long-timescale integration and abstract control, and posterior regions supporting immediate action selection (Badre & D'Esposito, 2009). Training and expertise are associated with greater engagement of anterior prefrontal regions and increased functional connectivity across temporal hierarchies (Koechlin et al., 2003; Nee & D'Esposito, 2016). The neuroscientific evidence suggests that *expertise enhances bidirectional coupling* between fast processing and slow structural control, a neural instantiation of P→S (monitoring) and S→P (control) dynamics.

Developmental studies show that children's metacognition is initially event-bound and reactive, becoming increasingly planful and temporally extended with age and instruction (Schneider, 2008). Young learners struggle to monitor learning over long intervals and to control behavior based on future outcomes; these abilities emerge gradually through adolescence, paralleling maturation of prefrontal control systems (Best & Miller, 2010).

## 7.5 Bidirectional Coupling Evidence and Limitations

Contemporary metacognition models emphasize that monitoring and control are reciprocally linked, not independent processes. Nelson and Narens' (1990, 1994) framework formalizes this bidirectionality as monitoring informing control decisions, and control altering future monitoring accuracy. Empirical work confirms that improved monitoring leads to better long-range control decisions, while effective control reshapes the informational basis for monitoring (Dunlosky & Metcalfe, 2009). This bidirectional loop aligns precisely with the claim and with DOT's P→S (monitoring) and S→P (control) structure.

Even experts show breakdowns when tasks exceed domain-specific temporal structure or when environments are highly volatile or adversarial. Under such conditions, monitoring collapses back to short-timescale cues, and control becomes reactive (Kahneman & Klein, 2009). This does not negate the claim but shows that temporal sophistication is domain- and structure-dependent. In highly automatized tasks, experts may monitor less consciously, relying on embodied or implicit control (Logan, 1988). However, when anomalies occur, experts can rapidly re-engage slow monitoring processes. This suggests flexible temporal scaling, not a loss of metacognition.

In summary, metacognitive monitoring becomes more temporally integrated with expertise, and control extends from reactive to prospective regulation. Bidirectional coupling between fast processing and slow structure strengthens with development and expertise, as indicated by neural control systems instantiating increasing temporal depth. Direct operationalization of "temporal span" in metacognitive monitoring and control needs further development, as does fine-grained measurement of P→S vs. S→P dynamics in real time.

These findings reframe metacognitive development as a fundamentally temporal achievement rather than a generic study-skills acquisition, and the reframing carries direct consequences for how teachers, curriculum designers, and educational leaders understand and support the development of self-regulation. A student who re-reads a paragraph immediately after failing to understand it, or who abandons a problem after a single unsuccessful attempt, or judges their own learning based on whether an idea felt familiar during the lesson, is not failing to apply good metacognitive habits. They are operating within a narrow temporal window in which fast, local signals dominate monitoring and control. From the perspective of this framework, such a student is not yet able to integrate performance information across the timescales needed for accurate self-assessment. Their processes-to-structures coupling is weak, and their structures-to-processes control and prediction capabilities remain reactive rather than prospective. Instruction that simply tells students to 'check their understanding' or 'reflect on their learning' without providing the temporal structure for doing so is asking novice learners to exercise a capability they have not yet developed and cannot develop through exhortation alone.

What this framework suggests instead is that metacognitive instruction must be scaffolded temporally, not just procedurally. A teacher who asks students to make a prediction before a lesson, revisit that prediction midway through, and then evaluate it at the end is not only running a comprehension exercise. They are extending the temporal span across which monitoring operates, creating the conditions for process-to-structure coupling to develop over minutes rather than seconds. A teacher who returns to a topic visited two weeks earlier and asks students to reconstruct what they understood then, compare it to what they understand now, and articulate what changed is extending that span further still, into the timescale at which slow attractor structures become visible rather than simply as 'what I know.' Delayed judgments of learning, iterative drafting with reflection between rounds, and longitudinal portfolio processes are not merely assessment strategies; they are interventions that develop the temporal architecture of metacognition by progressively demanding integration across wider spans. Crucially, the evidence reviewed above suggests that this development is specific to each domain. A student may have well-developed temporal monitoring in mathematics while remaining event-bound and reactive in historical analysis, because the slow attractor structures that anchor long-range monitoring must be sculpted within a domain, and cannot be transferred wholesale from one domain to another. (There may be a case to be made for generic metacognition if a domain shares a significant portion of the 'embodiment-to-symbolization' pathway with another domain. This debate is left for another day and article).

The leadership implications follow directly. If temporal sophistication in monitoring and control is domain-specific and structure-dependent, then developing it requires deliberate instructional design within each subject area, sustained across enough time for slow attractor landscapes (expert mental models, embedded professional development routines, supportive policies and processes of the organization) to form. This places a specific demand on curriculum leadership to ensure that sequences of lessons within a subject are designed with enough temporal depth and revisitation that students have repeated opportunities to monitor across intervals, not merely within them. It also reframes the persistent puzzle of why metacognitive interventions often show strong effects in controlled studies but weak transfer in institutional implementations. The answer, on this account, is not fidelity of delivery but temporal shallowness. Interventions that last only days or a few weeks cannot sculpt the slow-timescale attractor structures that robust metacognitive control requires. A reflective educational leader might ask: *In our current curriculum design, are students ever asked to monitor and regulate their own learning across timescales longer than a single lesson or unit, and if not, what structural changes would make that possible?*

## 8. Feedback topology: Research support

*Shortcuts in information processing (outputs direct to inputs without feedback, environment absorbing outputs as feedback) should map to specific temporal coupling patterns. For example, a person utilizing more shortcuts should exhibit richer multi-timescale integration. Such a person should have an increased ability to reshape environmental attractors at slower and longer timescales (e.g. should be able to manage more complex innovation processes).*

### 8.1 External Shortcuts as Distributed Feedback Loops

Research in distributed cognition demonstrates that cognition is not confined to the individual but is realized through feedback loops spanning brain, body, artifacts, and environment (Hutchins, 1995; Hollan et al., 2000). External artifacts (e.g. notes, diagrams, models, simulations, and dashboards) function as shortcuts that bypass internal memory limitations, stabilize information over time, and enable recursive interaction between internal and external states. These shortcuts effectively create additional feedback channels, allowing cognitive processes to operate simultaneously across multiple temporal scales. The evidence is mounting that more shortcuts correspond to a richer feedback topology, enabling coupling between fast internal cognition and slow environmental dynamics.

### 8.2 Cognitive Offloading and Multi-Timescale Integration

Cognitive offloading research shows that individuals who strategically use external resources do not merely reduce load, they expand the temporal horizon of cognition. Risko and Gilbert (2016) demonstrate, for example, that offloading allows agents to maintain long-term goals, track delayed consequences, and coordinate extended action sequences. Importantly, offloading supports recursive interaction. Outputs are externalized, persist, and later re-enter cognition as inputs, forming slow feedback loops absent in purely internal processing. Thus, external shortcuts enable integration across fast (seconds), medium (minutes), and slow (hours to months) timescales.

## 8.3 Epistemic Actions and Environmental Restructuring

Kirsh and Maglio's (1994) concept of *epistemic actions* provides direct support for the claim that agents reshape their environment to simplify cognition. These authors distinguish pragmatic actions, performed to bring one physically closer to a goal, from epistemic actions, performed to uncover information that is hidden or hard to mentally compute. Kirsh (2010) relatedly shows that such actions are not merely local optimizations; they alter the topology of the problem space, effectively reshaping environmental attractors, much as Epistemic Network Analysis alters the topology of semantic relationships in a dataset (Shaffer, 2016). The evidence strongly implies that repeated shortcut use leads to structural changes in the environment that operate at slower timescales than individual cognitive acts. Importantly, this provides an explicit mechanism for repeated real-time processing to become recalled control structures, which we refer to here as 'attractor reshaping.'

## 8.4 Innovation as Slow-Timescale Attractor Reshaping

Research on innovation and creative cognition emphasizes that innovation is fundamentally a slow-timescale, path-dependent process. Complex innovation requires iterative externalization (sketches, prototypes, models), persistence over long feedback delays, and accumulation and recombination of partial solutions into new holistic entities. Goel (1995) and Dunbar (1997) show that expert innovators rely heavily on external representations to explore and stabilize evolving solution spaces over time. Similarly, organizational learning research shows that artifacts (design documents, roadmaps, shared models) function as collective memory structures, shaping future innovation trajectories (Argote & Miron-Spektor, 2011), a feature referred to by Gibson and Ifenthaler (2025) as homologous functions across micro-meso and macro levels of a complex system. Thus, the ability to reshape environmental attractors is mediated by persistent external structures such as the "shortcuts" in the DOT taxonomy.

## 8.5Multi-Level Feedback in Complex Adaptive Systems

Complex systems research shows that systems capable of innovation and adaptation require multiple feedback loops operating at different timescales and mechanisms for amplifying small local changes into global reorganization (Holland, 1995; Levin, 1998). Human agents embedded in rich artifact ecologies effectively act as controllers of slow variables, adjusting environmental constraints that then shape fast behavior. Based on the observation that individuals who can create and exploit more feedback channels are better positioned to intervene at slow, structural levels of a system, we can project that an emerging expertise within an individual is better positioned to leverage structural levels to organize faster processing levels during performance.

## 8.6 Shortcut Density Links to Expertise and Performance

Expertise studies consistently show that experts externalize more information, use more intermediate representations, and construct layered representational ecologies. In scientific reasoning, for example, experts generate diagrams, annotations, and models that novices do not (Chi et al., 1988; Nersessian, 2008). These representations persist and evolve, enabling long-range coordination and innovation. Thus, shortcut richness correlates with the temporal depth semantic scope, and flexibility of cognitive control of an expert.

At the group level, teams with richer feedback infrastructures (shared artifacts, rapid prototyping, iterative review cycles) outperform others on complex, innovative tasks (Edmondson, 1999; Bechky, 2003). These infrastructures function as distributed shortcuts, allowing the organization to operate across multiple temporal scales simultaneously. We hold that homologous functions exist for distributed cognitive capabilities within an individual, leading to expertise that is better able to juggle multiple priorities during performance.

## 8.7 Feedback Topology Limitations

Research cautions that poorly designed artifacts can introduce noise, and excessive externalization can fragment both attention and motivation (Scaife & Rogers, 1996). Thus, shortcut *quality and alignment* matter as much as quantity. In addition, the relationship between shortcuts and innovation is strongest in ill-structured domains as well as design, science, engineering, and policy contexts. Less benefit is observed in tightly constrained or highly automated domains (habituated behaviors, simple problems), where fast thinking can work semi-autonomously.

In summary, the claim of a multi-leveled feedback topology is supported by research-based findings, including: 1) external shortcuts create additional feedback loops, 2) richer feedback topology enables multi-timescale integration, 3) persistent artifacts allow intervention at slow environmental timescales, and 4) innovation depends on the ability to reshape environmental attractors. Formal mapping between specific shortcut types and temporal coupling patterns, and quantitative measures of 'feedback topology richness' remain open areas for future research.

Translated into educational terms, feedback topology is not a property of assessment design alone; it is a structural feature of the entire learning environment that can be deliberately enriched or inadvertently impoverished at every level, from individual student practice to institutional design. At the student level, the difference between a learner who keeps a running problem-solving journal, annotates their drafts, maintains a concept map that evolves across a unit, and returns periodically to earlier work, and a learner who processes each task in isolation and discards it, illustrates a difference in feedback topology. The first student has constructed an external representational ecology that creates slow feedback loops; their past outputs persist, re-enter their thinking as inputs at later points, and gradually reshape their understanding. The second student, however capable in the moment, is operating with a temporally shallow feedback architecture in which only fast loops are available. Instruction that develops 'shortcut or feedback elaboration richness' (teaching students not merely to complete tasks but to externalize, annotate, revisit, and recombine their own cognitive products) develops a person's feedback infrastructure through which slow-timescale attractor sculpting becomes possible. This is more than a study-skills intervention; it is a structural intervention increasing the temporal depth of learning.

At the classroom level, a teacher's built-up collection of representational tools (such as worked examples left on the board, an evolving class concept map on the wall, annotated student responses used as discussion anchors, and the recurring problem types that advance and reinforce disciplinary thinking over weeks) forms the feedback structure of the instructional environment itself. A richly organized classroom, in this sense, is one where artifacts last long enough and are connected closely enough to serve as slow-timescale feedback channels, shaping the attractor landscape where daily fast interactions take place. A classroom without persistent representational artifacts, for example, where each lesson starts fresh, is characterized by a shallow feedback structure, regardless of how well-designed individual lessons are. The key point here is that it's not the number of artifacts that counts, but their quality and how well they align. A classroom filled with decorative displays that have no recursive connection from current to long-term learning tasks may be high in artifact quantity but lacks a rich feedback structure.

The institutional level is where the homologous functions claim becomes most consequential for educational leadership. A school that invests in curriculum mapping tools, shared assessment rubrics, longitudinal student portfolios, data systems that track learning trajectories across years, and professional learning communities in which teachers iteratively review and revise their practice is constructing an organizational feedback topology that operates at the slowest timescales. Such a system allows local, fast-moving classroom interactions to accumulate into structural changes in the institution's attractor landscape over time. Conversely, a school whose improvement efforts consist of discrete, non-connected initiatives has a shallow organizational feedback topology regardless of the effort invested in individual components. Which would be the case if each year's professional development replaces rather than builds on the last, assessment data is collected but not recursively used, and curriculum frameworks are revised without reference to what the previous framework revealed.

The practical implication is that educational leaders should evaluate their institutional design not only by the quality of individual feedback mechanisms but by the degree to which those mechanisms are coupled across timescales. Leaders need to help ensure that fast-loop classroom data informs medium-loop unit revision, which in turn reshapes slow-loop curriculum and professional learning architecture. A reflective educational leader might ask: *What persistent external structures in our school allow the outputs of teaching and learning today to re-enter and reshape the conditions for teaching and learning next month, next year, and across the span of a student's time with us, and where are those feedback loops broken or missing entirely?*

# 9 Schema flexibility paradox: Research support

*Cognitive load theory's 'expertise reversal effect' occurs when highly compressed attractors (expert schemas) cannot flexibly respond to novelty, which is explained by over-optimization for specific temporal predictions.*

## 9.1 The Expertise Reversal Effect in Cognitive Load Theory

The expertise reversal effect is well established in cognitive load theory: instructional supports that benefit novices (worked examples, explicit guidance) become ineffective or harmful for experts (Kalyuga et al., 2003; Kalyuga, 2007; Sweller et al., 2019). Empirically, experts experience increased extraneous load from redundant information and performance decrements when instruction conflicts with existing schemas. This implies that expert schemas are highly compressed representations that are efficient under familiar conditions but fragile under mismatch. These considerations lead to the conclusion that the expertise reversal effect reflects not just redundancy, but *misalignment between compressed representations and task structure*.

## 9.2 Schema Compression and Loss of Flexibility

Expertise research shows that experts rely on pattern-based, chunked representations that drastically reduce processing demands (Chase & Simon, 1973; Gobet et al., 2001). However, these representations encode regularities at the expense of atypical features and are tuned to expected task distributions. When a problem's structure deviates, experts may misclassify novel cases, perseverate on familiar strategies, or exhibit slower adaptation than advanced novices (Chi et al., 1988; Dane, 2010). Based on this evidence, *schema compression trades flexibility for efficiency*, which is consistent with an "over-optimized attractor" account.

## 9.3 Cognitive Flexibility and Negative Transfer

Research on negative transfer shows that prior expertise can hinder learning in structurally similar but functionally different domains (Barnett & Ceci, 2002; Day & Goldstone, 2012). Goldstone and colleagues, for example, show that when learners internalize highly specific perceptual or conceptual encodings, they struggle to re-represent problems under new constraints. If schemas are domain-specific multiscale attractors rather than portable symbolic structures, as advocated here, then transfer is not the application of a representation to a new context but the partial reuse of an attractor landscape under new boundary conditions, a much harder, slower, and more constrained process than most transfer literature assumes. This is especially pronounced when representations have become automatized and implicit, limiting access for reconfiguration. This implies that *over-optimized schemas resist restructuring when temporal or structural contingencies change*.

## 9.4 Neuroscientific Evidence for Predictive Over-Optimization

Predictive processing frameworks emphasize that learning minimizes prediction error by tuning internal models to expected input distributions. However, overly precise predictions reduce sensitivity to novel signals (Friston, 2010). Neuroimaging studies show that experts exhibit reduced neural variability and stronger, faster top-down predictions, but at the same time, they exhibit greater prediction error when expectations are violated and are slower in updating predictions under unexpected conditions (Garrett et al., 2013; Summerfield & de Lange, 2014). From this evidence, we understand that expert models are temporally optimized for expected environments; *novelty induces maladaptive prediction error cascades*, which is an attractor rigidity effect.

## 9.5 Temporal Specialization in Expertise

Expert schemas operate at longer temporal scales, enabling rapid anticipation and planning (Ericsson & Kintsch, 1995). However, this temporal depth can impair responsiveness when short-term contingencies change abruptly. Kahneman and Klein (2009) show, for example, that *intuitive expertise breaks down in environments with low regularity and delayed or noisy* *feedback.* These are precisely the conditions where long-term temporal predictions are unreliable.

### 9.6 Constraints and Limitations of Schema Flexibility

Not all experts exhibit rigidity or over-optimization. Experts trained in ill-structured domains (e.g. the arts, design) and those who work in highly variable or adversarial environments (e.g., emergency medicine, elite sports) show greater adaptability (Ericsson, 2006). This evidence suggests that *rigidity emerges when expertise develops under stable temporal regimes, reinforcing narrow attractors*. Experts with strong metacognitive control can detect schema failure and deliberately slow processing to reconfigure representations (Dunlosky & Metcalfe, 2009). This supports the claim that temporal coupling sophistication mediates flexibility.

Cognitive load theory (CLT) traditionally explains the expertise reversal effect via redundancy effects. The dynamic attractor formulation goes further by specifying *why* redundancy is harmful: redundant information forces experts to reconcile slow, compressed predictions with fast, conflicting inputs, and this creates cross-timescale interference, increasing extraneous load. Thus, the schema flexibility paradox reframes the expertise reversal effect as a temporal mismatch problem, not merely an instructional design issue.

Over-optimization is a continuous risk at all developmental levels, not a unitary stage to be passed through once resolved. So, the conditions that prevent over-optimization (structured variability, metacognitive temporal reach, persistent external representations, and exposure to genuinely novel demands) need to be built into learning environments. A learning environment that treats experts and students with high capabilities as finished products and only treats novices as 'works in progress' has the developmental logic exactly backwards from what this framework suggests.

In summary, the claim of schema flexibility via dynamic temporal hierarchical mechanisms is supported by evidence that expert schemas are highly compressed and efficient, compression reduces flexibility under novelty, expertise reversal reflects misalignment between instruction and optimized schemas, and predictive over-optimization explains rigidity under temporal mismatch. Direct measurement of 'attractor rigidity' and experimental manipulation of temporal prediction horizons in experts remain issues for future research and theoretical development.

## Open Questions for Future Research

The account offered here, despite its complexity, is incomplete. Learners clearly differ in how they couple fast and slow processes. They differ, for example, by neurological profile, developmental stage, cultural background, and prior experience, and we have not dealt with these differences here. A second missing piece is the role of emotion and motivation. Attractor sculpting at slow timescales is profoundly shaped by whether learners or professionals are in a state of psychological safety, whether they have reasons to persist through the high-cost apprentice stage, and whether the institutional environment reinforces or punishes the vulnerability that genuine accommodation requires. A third missing piece is a measurement framework at the classroom level. While we propose empirical tests in the appendix, none of them operationalize the core constructs of temporal depth or attractor richness and flexibility for classroom practitioners. But we hope to have made new ideas in psychological modeling, field research, and practical applications in education salient for further reflection and study.

The reconceptualization of representation as multiscale attractors opens several novel research directions. Methodologically, new empirical techniques are needed that distinguish prediction error across timescales and spatial hierarchies rather than collapsing cognitive phenomena into undifferentiated 'cognitive load' measures. At the individual level, learners may vary in their optimal timescale separation, perhaps explaining why some thrive with verbal scaffolding that unfolds over time while others prefer spatial representations that remain persistently available for embodied interaction. Cross-culturally, different learning practices may establish distinct hierarchical prediction architectures, offering a dynamic systems explanation for observed variations in metacognitive strategies across cultures.

Finally, the design of digital learning environments takes on new significance. Interfaces can either disrupt natural timescale hierarchies or enhance them, transforming the external topology derived from the DOT framework from a descriptive frame into a prescriptive design principle for simulation, further theory development, and field-based research. Understanding learning as multiscale attractor dynamics shifts questions like 'how much information can working memory hold?' to questions like 'how do we design adaptive learning environments that support prediction across coupled timescales and hierarchical systems of organization?'

Perhaps the single most important takeaway from this article for educators is the need to *design for temporal depth and hierarchical integration*, not just 'content coverage.' Every major implication in the article (embodied scaffolding develops before symbolic compression, metacognitive instruction must be structured across longer intervals, feedback needs to be enriched across timescales, and students as well as professionals learn by building on strengths rather than deficits) reduces to the same underlying principle: *the temporal architecture of a learning environment determines what kinds of cognitive structures can form within it*. Many current educational designs operate at timescales that are too shallow to support the slow attractor sculpting required for durable learning, flexible expertise, and genuine metacognitive development. Leaders and teachers should therefore redesign learning environments for continuity over longer timescales of months to years (not one lesson at a time, with a quiz at the end of the class) to ensure that outputs, feedback, and the lessons learned at each timescale persist long enough to be utilized as inputs for further learning.

The reconceptualization of *representations as dynamic multiscale attractors* doesn't eliminate all tensions in cognitive science theories but hopefully may transform them into a reconciliation of complementary perspectives. The new perspective discussed here has included cognitive load theory's description of compression requirements, embodied cognition's description of distributed implementation, and the multiscale attractor framework's formal taxonomy of trajectories of expertise for analyzing the integration of a variety of cognitive and metacognitive mechanisms across stages of development.

## References

Argote, L., & Miron-Spektor, E. (2011). Organizational learning: From experience to knowledge. *Organization Science, 22*(5), 1123–1137. https://doi.org/10.1287/orsc.1100.0621

Ayres, P. (2006). Using subjective measures to detect variations of intrinsic cognitive load within problems. *Learning and Instruction, 16*(5), 389–400. https://doi.org/10.1016/j.learninstruc.2006.09.001

Baber, C., Howes, A., & Cross, E. S. (2013). Sensori-motor representations and cognitive offloading for tool use. *Human–Computer Interaction, 28*(1), 1–36. https://doi.org/10.1080/07370024.2012.697500

Badre, D., & D'Esposito, M. (2009). Is the rostro-caudal axis of the frontal lobe hierarchical? *Nature Reviews Neuroscience, 10*(9), 659–669. https://doi.org/10.1038/nrn2667

Baum, E. (2003). What is thought. MIT.

Barnett, S. M., & Ceci, S. J. (2002). When and where do we apply what we learn? *Psychological Bulletin, 128*(4), 612–637. https://doi.org/10.1037/0033-2909.128.4.612

Bar-Yam, Y. (1997). General Features of Complex Systems. *Knowledge Management, Organisational Intelligence and Learning and Complexity*, *I*.

Bechky, B. A. (2003). Sharing meaning across occupational communities: The transformation of understanding on a production floor. *Organization Science, 14*(3), 312–330. https://doi.org/10.1287/orsc.14.3.312.15162

Best, J. R., & Miller, P. H. (2010). A developmental perspective on executive function. *Child Development, 81*(6), 1641–1660. https://doi.org/10.1111/j.1467-8624.2010.01499.x

Brooks, R. A. (1991). Intelligence without representation. Artificial Intelligence, 47, 139–159. https://doi.org/10.1016/0004-3702(91)90053-M

Chandler, P., & Sweller, J. (1991). Cognitive load theory and the format of instruction. *Cognition and Instruction, 8*(4), 293–332. https://doi.org/10.1207/s1532690xci0804_2

Chase, W. G., & Simon, H. A. (1973). Perception in chess. *Cognitive Psychology, 4*(1), 55–81. https://doi.org/10.1016/0010-0285(73)90004-2

Chi, M. T. H., Glaser, R., & Farr, M. J. (1988). *The nature of expertise*. Erlbaum.

Cook, S. W., Yip, T. K., & Goldin-Meadow, S. (2012). Gesture makes memories that last. *Journal of Memory and Language, 66*(4), 661–674. https://doi.org/10.1016/j.jml.2012.04.007

Coull, J. T., Cheng, R. K., & Meck, W. H. (2013). Neuroanatomical and neurochemical substrates of timing. *Neuropsychopharmacology, 36*(1), 3–25. https://doi.org/10.1038/npp.2010.113

Dane, E. (2010). Reconsidering the trade-off between expertise and flexibility. *Academy of Management Review, 35*(4), 579–603. https://doi.org/10.5465/AMR.2010.53502832

Day, S. B., & Goldstone, R. L. (2012). The import of knowledge export: Connecting findings and theories of transfer of learning. *Educational Psychologist, 47*(3), 153–176. https://doi.org/10.1080/00461520.2012.696438

Dunbar, K. (1997). How scientists think: Online creativity and conceptual change in science. *Cognitive Science, 21*(4), 397–434. https://doi.org/10.1207/s15516709cog2104_2

Dunlosky, J., & Metcalfe, J. (2009). *Metacognition*. Sage.

Edmondson, A. (1999). Psychological safety and learning behavior in work teams. *Administrative Science Quarterly, 44*(2), 350–383. https://doi.org/10.2307/2666999

Efklides, A. (2011). Interactions of metacognition with motivation and affect in self-regulated learning. *Educational Psychologist, 46*(1), 6–25. https://doi.org/10.1080/00461520.2011.538645

Ericsson, K. A. (2006). The influence of experience and deliberate practice on the development of superior expert performance. *Cambridge Handbook of Expertise and Expert Performance*, 683–703.

Ericsson, K. A., & Kintsch, W. (1995). Long-term working memory. *Psychological Review, 102*(2), 211–245. https://doi.org/10.1037/0033-295X.102.2.211

Friston, K. (2010). The free-energy principle: A unified brain theory? *Nature Reviews Neuroscience, 11*(2), 127–138. https://doi.org/10.1038/nrn2787

Garrett, D. D., Samanez-Larkin, G. R., MacDonald, S. W. S., Lindenberger, U., McIntosh, A. R., & Grady, C. L. (2013). Moment-to-moment brain signal variability: A next frontier in human brain mapping? *Neuroscience & Biobehavioral Reviews, 37*(4), 610–624. https://doi.org/10.1016/j.neubiorev.2013.02.015

Gell-mann, M. (1995). What is complexity. In Complexity (Vol. 1). Santa Fe Institute.

Gibson, D., & Azukas, E. (2025). A Taxonomy of Metacognitive Learning Scenarios in Professional Contexts: Integrating Systems Theory with Empirical Constraints. In Preparation. https://docs.google.com/document/d/1lmEDPZLNDf7Vt4BkSiWdQBMeS5lvwr6KmRK_ba2J7yY/edit?usp=sharing

Gigerenzer, G., & Todd, P. (1999a). Simple heuristics that make us smart. Oxford University Press.

Gigerenzer, G., & Todd, P. M. (1999b). Fast and frugal heuristics: The adaptive toolbox. In Simple heuristics that make us smart (pp. 3–34). https://doi.org/10.1002/mar.10060

Gobet, F., Lane, P. C. R., Croker, S., Cheng, P. C. H., Jones, G., Oliver, I., & Pine, J. M. (2001). Chunking mechanisms in human learning. *Trends in Cognitive Sciences, 5*(6), 236–243. https://doi.org/10.1016/S1364-6613(00)01662-4

Goel, V. (1995). *Sketches of thought*. MIT Press.

Goldin-Meadow, S., Cook, S. W., & Mitchell, Z. A. (2007). Gesturing gives children new ideas about math. *Psychological Science, 18*(6), 507–513. https://doi.org/10.1111/j.1467-9280.2007.01945.x

Goldin-Meadow, S., Nusbaum, H., Kelly, S. D., & Wagner, S. (2001). Explaining math: Gesturing lightens the load. *Psychological Science, 12*(6), 516–522. https://doi.org/10.1111/1467-9280.00395

Hasson, U., Yang, E., Vallines, I., Heeger, D. J., & Rubin, N. (2008). A hierarchy of temporal receptive windows in human cortex. *Journal of Neuroscience, 28*(10), 2539–2550. https://doi.org/10.1523/JNEUROSCI.5487-07.2008

Hattie, J., & Temperley, H. (2007). The power of feedback. Review of Educational Research, 77(1), 81–112.

Hawkins, J. (2021). A thousand brains: A new theory of intelligence. Basic Books.

Hawkins, J., & Blakeslee, S. (2004). On intelligence. Henry Holt and Company.

Holland, J. (1998). *Emergence: From chaos to order*. In Reading, MA: Helix Books Addison Wesley. Perseus Books Group.

Holland, J. (2019). *Complex adaptive systems: A primer*. In D. Krakauer (Ed.), Worlds hidden in plain sight: Thirty years of complexity thinking at the Santa Fe Institute. Santa Fe Institute.

Holland, J., Hutchins, E., & Kirsh, D. (2000). Distributed cognition: Toward a new foundation for human–computer interaction research. *ACM Transactions on Computer–Human Interaction, 7*(2), 174–196. https://doi.org/10.1145/353485.353487

Holland, J. H. (1995). *Hidden order: How adaptation builds complexity*. Addison-Wesley.

Hutchins, E. (1995). *Cognition in the wild*. MIT Press.

Ifenthaler, D. (2011). Bridging the Gap between Expert-Novice Differences: The Model-Based Feedback Approach. Journal of Research on Technology in Education, 43(2), 103–117.

Johnson, T. E., O'Connor, D. L., Pirnay-Dummer, P. N., Ifenthaler, D., Spector, J. M., & Seel, N. (2006). Comparative Study of Mental Model Research Methods: Relationships Among ACSMM, SMD, MITOCAR & DEEP Methodologies. In A. J. Cañas & J. D. Novak (Eds.), Concept Maps: Theory, Methodology, Technology. Proc. Of the Second Int. Conference on Concept Mapping. (Vol. 1, pp. 87–94). Universidad de Costa Rica. http://eprint.ihmc.us/164/

Johnson-Glenberg, M. C., Birchfield, D. A., Tolentino, L., & Koziupa, T. (2014). Collaborative embodied learning in mixed reality motion-capture environments. *Educational Psychology Review, 26*(3), 463–491. https://doi.org/10.1007/s10648-014-9253-2

Kahneman, D. (2011). Thinking fast and thinking slow. In Farrar, Strauss and Giroux, New York, NY.

Kahneman, D., & Klein, G. (2009). Conditions for intuitive expertise: A failure to disagree. *American Psychologist, 64*(6), 515–526. https://doi.org/10.1037/a0016755

Kalyuga, S. (2007). Expertise reversal effect and its implications for learner-tailored instruction. *Educational Psychology Review, 19*(4), 509–539. https://doi.org/10.1007/s10648-007-9054-3

Kalyuga, S., Ayres, P., Chandler, P., & Sweller, J. (2003). The expertise reversal effect. *Educational Psychologist, 38*(1), 23–31. https://doi.org/10.1207/S15326985EP3801_4

Kauffman, S. A. (1993). *The origins of order: Self-organization and selection in evolution*. Oxford University Press.

Kelso, J. A. S. (1995). *Dynamic patterns: The self-organization of brain and behavior*. MIT Press.

Kelso, J. A. S., & Engstrøm, D. A. (2006). *The complementary nature*. MIT Press.

Kirsh, D. (2010). Thinking with external representations. *AI & Society, 25*(4), 441–454. https://doi.org/10.1007/s00146-010-0272-8

Kirsh, D., & Maglio, P. (1994). On distinguishing epistemic from pragmatic action. *Cognitive Science, 18*(4), 513–549. https://doi.org/10.1207/s15516709cog1804_1

Koechlin, E., Ody, C., & Kouneiher, F. (2003). The architecture of cognitive control in the human prefrontal cortex. *Science, 302*(5648), 1181–1185. https://doi.org/10.1126/science.1088545

Koriat, A. (1997). Monitoring one's own knowledge during study: A cue-utilization approach. *Journal of Experimental Psychology: General, 126*(4), 349–370. https://doi.org/10.1037/0096-3445.126.4.349

Langer, S. K. (1942). Philosophy in a new key; a study in the symbolism of reason rite and art. Harvard University Press.

Leppink, J., Paas, F., van Gog, T., van der Vleuten, C. P. M., & van Merriënboer, J. J. G. (2013). Effects of pairs of problems and examples on task performance and different types of cognitive load. *Learning and Instruction, 30*, 32–42. https://doi.org/10.1016/j.learninstruc.2013.12.001

Lerner, Y., Honey, C. J., Silbert, L. J., & Hasson, U. (2011). Topographic mapping of a hierarchy of temporal receptive windows using a narrated story. *Journal of Neuroscience, 31*(8), 2906–2915. https://doi.org/10.1523/JNEUROSCI.3684-10.2011

Levin, S. A. (1998). Ecosystems and the biosphere as complex adaptive systems. *Ecosystems, 1*(5), 431–436. https://doi.org/10.1007/s100219900037

Logan, G. D. (1988). Toward an instance theory of automatization. *Psychological Review, 95*(4), 492–527. https://doi.org/10.1037/0033-295X.95.4.492

Nee, D. E., & D'Esposito, M. (2016). The hierarchical organization of the lateral prefrontal cortex. *eLife, 5*, e12112. https://doi.org/10.7554/eLife.12112

Nelson, T. O., & Narens, L. (1990). Metamemory: A theoretical framework and new findings. *Psychology of Learning and Motivation, 26*, 125–173. https://doi.org/10.1016/S0079-7421(08)60053-5

Nelson, T. O., & Narens, L. (1994). Why investigate metacognition? *Metacognition: Knowing about knowing*, 1–25.

Nersessian, N. J. (2008). *Creating scientific concepts*. MIT Press.

Newell, A. (1990). *Unified theories of cognition*. Harvard University Press.

Pashler, H. (1994). Dual-task interference in simple tasks: Data and theory. *Psychological Bulletin, 116*(2), 220–244. https://doi.org/10.1037/0033-2909.116.2.220

Piaget, J. (1985). The equilibration of cognitive structures: The central problem of intellectual development. In Chicago: University of Chicago Press.

Prigogine, I., & Stengers, I. (1984). *Order out of chaos: Man's new dialogue with nature*. Bantam Books.

Reimer, T., Raeskamp, J., Rieskamp, J., & Raeskamp, J. (2006). Fast and frugal heuristics. In Encyclopedia of Social Psychology (Vol. 7). https://doi.org/10.1016/j.psychsport.2006.06.002

Rhodes, M. G., & Tauber, S. K. (2011). The influence of delaying judgments of learning on metacognitive accuracy. *Journal of Experimental Psychology: Learning, Memory, and Cognition, 37*(2), 410–424. https://doi.org/10.1037/a0021705

Risko, E. F., & Gilbert, S. J. (2016). Cognitive offloading. *Trends in Cognitive Sciences, 20*(9), 676–688. https://doi.org/10.1016/j.tics.2016.07.002

Rubinov, M., & Sporns, O. (2010). Complex network measures of brain connectivity: Uses and interpretations. NeuroImage. https://doi.org/10.1016/j.neuroimage.2009.10.003

Scaife, M., & Rogers, Y. (1996). External cognition: How do graphical representations work? *International Journal of Human–Computer Studies, 45*(2), 185–213. https://doi.org/10.1006/ijhc.1996.0048

Schneider, W. (2008). The development of metacognitive knowledge in children and adolescents. *Developmental Review, 28*(3), 229–256. https://doi.org/10.1016/j.dr.2007.12.001

Schöner, G. (2008). Dynamical systems approaches to cognition. In R. Sun (Ed.), *The Cambridge handbook of computational psychology* (pp. 101–126). Cambridge University Press.

Shaffer, D. (2016). A Tutorial on Epistemic Network Analysis. *Journal of Learning Analytics*, *3*(3), 9–45.

Simon, H. A. (1962). The architecture of complexity. *Proceedings of the American Philosophical Society*, *106*(6), 467–482.

Skulmowski, A., & Rey, G. D. (2018). Embodied learning: Introducing a taxonomy based on bodily engagement and task integration. *Cognitive Research: Principles and Implications, 3*(6). https://doi.org/10.1186/s41235-018-0092-9

Strogatz, S. H. (2015). *Nonlinear dynamics and chaos: With applications to physics, biology, chemistry, and engineering* (2nd ed.). Westview Press.

Summerfield, C., & de Lange, F. P. (2014). Expectation in perceptual decision making: Neural and computational mechanisms. *Nature Reviews Neuroscience, 15*(11), 745–756. https://doi.org/10.1038/nrn3838

Sweller, J. (2010). Element interactivity and intrinsic, extraneous, and germane cognitive load. *Educational Psychology Review, 22*(2), 123–138. https://doi.org/10.1007/s10648-010-9128-5

Sweller, J., Ayres, P., & Kalyuga, S. (2019). *Cognitive load theory* (2nd ed.). Springer. https://doi.org/10.1007/978-3-319-91988-9

Thagard, P. (2005). Abductive inference: From philosophical analysis to neural mechanisms.

Thagard, P. (2010). How Brains Make Mental Models. In L. Magnani, W. Carnielli, & C. Pizzi (Eds.), Model-Based Reasoning in Science and Technology (Vol. 314, pp. 447–461). Springer Berlin Heidelberg. https://doi.org/10.1007/978-3-642-15223-8_25

Thelen, E., & Smith, L. B. (1994). *A dynamic systems approach to the development of cognition and action*. MIT Press.

Tombu, M., & Jolicoeur, P. (2003). A central capacity sharing model of dual-task performance. *Journal of Experimental Psychology: Human Perception and Performance, 29*(1), 3–18. https://doi.org/10.1037/0096-1523.29.1.3

Unsworth, N., & Robison, M. K. (2015). Individual differences in the allocation of attention to items in working memory. *Journal of Experimental Psychology: Learning, Memory, and Cognition, 41*(4), 1057–1076. https://doi.org/10.1037/xlm0000074

van der Wel, P., & van Steenbergen, H. (2018). Pupil dilation as an index of effort in cognitive control tasks. *Psychonomic Bulletin & Review, 25*(6), 2005–2015. https://doi.org/10.3758/s13423-018-1432-y

Van Orden, G. C., Holden, J. G., & Turvey, M. T. (2003). Self-organization of cognitive performance. *Journal of Experimental Psychology: General*, *132*(3), 331–350. https://doi.org/10.1037/0096-3445.132.3.331

Varela, F. J., Thompson, E., & Rosch, E. (1991). *The embodied mind: Cognitive science and human experience*. MIT Press.

Wilson, A. D., & Golonka, S. (2013). Embodied cognition is not what you think it is. *Frontiers in Psychology, 4*, 58. https://doi.org/10.3389/fpsyg.2013.00058

Winne, P. H., & Hadwin, A. F. (2008). The weave of motivation and self-regulated learning. *Motivation and Self-Regulated Learning*, 297–314. https://doi.org/10.4324/9780203831076

Wolpert, D. M., & Ghahramani, Z. (2000). Computational principles of movement neuroscience. *Nature Neuroscience, 3*(11), 1212–1217. https://doi.org/10.1038/81497

Xiong, M. (2022). Artificial intelligence and causal inference (First edition). CRC Press, Taylor & Francis Group.

Zimmerman, B. J. (2002). Becoming a self-regulated learner: An overview. *Theory Into Practice, 41*(2), 64–70. https://doi.org/10.1207/s15430421tip4102_2

Zylberberg, A., Fernandez Slezak, D., Roelfsema, P. R., Dehaene, S., & Sigman, M. (2010). The human Turing machine: A neural framework for mental programs. *Trends in Cognitive Sciences, 15*(7), 293–300. https://doi.org/10.1016/j.tics.2011.05.007

# APPENDIX 1: Proposed empirical studies examining cross-timescale interference.

## 5.1 Cognitive Load, Element Interactivity, and Temporal Coordination

**Hypothesis:** Learners will exhibit disproportionately higher cognitive load when task elements must be coordinated across incompatible temporal scales than when the same number of elements can be nested hierarchically.

**Setting:** A computer-based learning environment in which participants solve multi-step science problems (e.g., interpreting a chemical reaction). In the *hierarchical* condition, fast-changing variables (e.g., molecular collision animations updating every second) are explicitly nested within slower explanatory structures (e.g., a stepwise procedural guide updating every 30 seconds). In the *non-hierarchical* condition, the same variables update independently at competing rates with no organizational nesting. Participants must track both streams simultaneously without temporal scaffolding.

**Subjects:** Undergraduate students (N ≈ 120) with comparable prior knowledge, randomly assigned to hierarchical or non-hierarchical conditions, with a possible third control condition presenting elements sequentially.

**Data Collection:** Intrinsic cognitive load is measured via a validated self-report instrument (e.g., Leppink et al., 2013), supplemented by a concurrent secondary task (e.g., responding to periodic auditory probes) to capture objective load. Problem-solving accuracy and time-on-task are recorded. Critically, element interactivity is held constant across conditions; only the temporal organization of those elements differs.

**Analysis:** Performance and load measures are compared across conditions using standard inferential methods. To capture the attractor dynamics central to the framework, participants' problem-solving trajectories are plotted in a state space defined by accuracy and response latency over successive task steps. Recurrence quantification analysis (RQA) is applied to these trajectories to determine whether the hierarchical condition produces more stable, periodic attractor patterns (indicating organized temporal compression) while the non-hierarchical condition produces more chaotic, less recurrent trajectories (indicating failed compression and elevated load). A significant interaction between temporal organization and load, beyond what element count alone would predict, would support the claim that temporal misalignment is an independent source of intrinsic load.

## 5.2 Dual-Task and Psychological Refractory Period (PRP) Evidence

**Hypothesis:** Dual-task interference will be greater when two concurrent tasks operate at temporally incompatible rates than when they operate at rates that permit hierarchical nesting, even when overall task difficulty is equated.

**Setting:** A classic PRP paradigm modified to manipulate temporal compatibility. Participants perform two tasks simultaneously: a fast task (e.g., responding to brief visual stimuli appearing every 500 ms) and a slow task (e.g., tracking a gradually shifting auditory tone that changes over 5-second intervals). In the *compatible* condition, slow-task changes occur at boundaries that align with natural groupings of fast-task events (e.g., every 10th stimulus), permitting hierarchical nesting. In the *incompatible* condition, slow-task changes occur at irregular intervals that cut across fast-task groupings, preventing nesting.

**Subjects:** Adults ($N \approx 80$), screened for normal vision and hearing, randomly assigned to compatible or incompatible conditions in a within-subjects crossover design with counterbalancing.

**Data Collection:** Reaction times and error rates for both tasks are recorded on every trial, along with inter-response intervals. Subjective load ratings are collected after each block.

**Analysis:** Beyond comparing mean reaction times and error rates across conditions, the temporal microstructure of responses is analyzed. Cross-recurrence quantification analysis (CRQA) between the fast-task and slow-task response streams assesses the degree to which the two behavioral timescales become coupled. In the compatible condition, CRQA should reveal structured coupling (the fast-task response pattern phase-locks to slow-task boundaries), indicating a stable multi-scale attractor. In the incompatible condition, coupling should be weak or erratic, reflecting competing attractor basins and manifesting as larger PRP effects. The key prediction is that interference magnitude correlates with the degree of cross-timescale decoupling, not simply with the number of concurrent tasks.

## 5.3 Hierarchical Temporal Processing in the Brain

**Hypothesis:** When learners must simultaneously track information at fast and slow timescales that cannot be hierarchically organized, frontoparietal control networks will show increased activation and decreased coupling with temporal-hierarchy regions, reflecting the neural cost of violated timescale separation.

**Setting:** An fMRI study in which participants watch narrated instructional videos. In the *hierarchically organized* condition, the narration has clear paragraph-level structure that nests sentence-level information within broader thematic arcs (e.g., a well-structured science lecture). In the *temporally disorganized* condition, the same sentences are rearranged so that local coherence (sentence-to-sentence) is preserved but long-range thematic structure is disrupted—participants must simultaneously process moment-to-moment content and attempt to extract an overarching structure that is not hierarchically available.

**Subjects:** Adults ($N \approx 40$) with no expertise in the lecture topic, screened for MRI compatibility.

**Data Collection:** Blood Oxygen Level Dependent (BOLD) signal is acquired continuously during both conditions. Post-scan comprehension tests assess learning at both local (sentence-level fact recall) and global (thematic integration) levels. Regions of interest include sensory cortices (fast temporal receptive windows), temporal association areas (medium windows), and prefrontal cortex (slow windows), following the hierarchy identified by Hasson et al. (2008).

**Analysis:** Inter-subject correlation (ISC) analysis assesses the reliability of neural responses at each level of the temporal hierarchy across conditions. The hierarchically organized condition should produce strong ISC across all levels, reflecting a stable multi-scale neural attractor in which slow regions constrain fast regions. The disorganized condition should preserve fast-level ISC (local sentence processing remains intact) but reduce slow-level ISC and increase frontoparietal activation, a neural signature of failed temporal nesting requiring effortful control. Dynamic functional connectivity analysis further tests whether the coupling between fast and slow cortical regions weakens in the disorganized condition, consistent with attractor decoupling across timescales.

## 5.4 Attention to Competing Temporal Features

**Hypothesis:** Attending to two temporal patterns simultaneously will produce greater error and slower responses when the patterns are temporally orthogonal than when one pattern hierarchically organizes the other.

**Setting:** A rhythm-monitoring task in which participants listen to auditory sequences containing two superimposed temporal structures: a fast beat (e.g., 4 Hz) and a slow modulation (e.g., 0.5 Hz amplitude envelope). In the *hierarchical* condition, the slow modulation groups the fast beats into perceptually natural clusters (e.g., strong-weak-weak-weak). In the *orthogonal* condition, the slow modulation cycle length is set to a non-integer ratio of the fast beat (e.g., every 3.7 beats), preventing clean nesting. Participants must detect occasional deviants in both the fast pattern (a missing beat) and the slow pattern (an unexpected amplitude shift).

**Subjects:** Musically untrained adults ($N \approx 60$), randomly assigned to conditions, with a within-subjects replication using counterbalanced ordering.

**Data Collection:** Detection accuracy and reaction time for deviants at each timescale, along with concurrent EEG recording to capture mismatch negativity (MMN) and P300 components associated with prediction error at fast and slow timescales respectively.

**Analysis:** Behaviorally, a timescale × compatibility interaction is predicted: detection at both scales should suffer more in the orthogonal condition than the hierarchical condition. The EEG data are analyzed using time-frequency decomposition to examine whether neural oscillations at the two target frequencies show phase-amplitude coupling (the slow oscillation's phase modulating the fast oscillation's amplitude) in the hierarchical condition, a neural marker of a coupled multi-scale attractor. In the orthogonal condition, this coupling should be absent or reduced, and frontal theta power (associated with cognitive control) should increase, providing a convergent neural indicator that the absence of temporal hierarchy imposes additional attentional cost.

## 5.5 Physiological Load Indicators

**Hypothesis:** Pupil dilation and neural load signatures will increase non-linearly as the number of temporally independent scales that must be simultaneously coordinated increases, exceeding the additive cost predicted by task difficulty alone.

**Setting:** A graded load paradigm in which participants perform an air-traffic-control simulation requiring monitoring of aircraft at progressively more temporal scales: one scale (current positions only), two scales (current positions plus estimated arrival sequences over the next minute), and three scales (current positions, arrival sequences, and a slowly evolving weather system that shifts flight paths over five minutes). Crucially, in each condition the total number of information elements is equated by adjusting the number of aircraft; only the number of independent temporal scales increases.

**Subjects:** Adults (N ≈ 50) with no aviation experience, tested in a within-subjects design with conditions presented in counterbalanced blocks.

**Data Collection:** Continuous pupillometry (tonic and phasic pupil diameter), task performance (collision avoidance accuracy, decision latency), and subjective cognitive load ratings after each block. Optional fNIRS over prefrontal cortex provides a secondary neural load measure.

**Analysis:** If temporal coordination is an independent load source, load indicators should show a non-linear (superadditive) increase from one to three timescales, even though element count is held constant. To connect this to attractor dynamics, the variability structure of pupil diameter time series is analyzed using detrended fluctuation analysis (DFA). Under low temporal-scale demands, pupil fluctuations should exhibit long-range correlations characteristic of a system operating near a stable attractor. As the number of uncoupled timescales increases, these correlations should break down. The scaling exponent should shift toward randomness—indicating that the cognitive system has been pushed away from organized multi-scale attractor dynamics and into a regime of effortful, unstable coordination.

## 5.6 Limitations: Expertise and Temporal Chunking

**Hypothesis:** Experts will maintain stable multi-scale attractor dynamics (and lower cognitive load) under cross-timescale demands that destabilize novice performance, and this advantage will be mediated by the experts' ability to restore hierarchical temporal organization through chunking.

**Setting:** A domain-specific problem-solving task (e.g., medical diagnosis from evolving patient data) in which participants must simultaneously track fast-changing vital signs (updated every few seconds) and slow-evolving lab results (updated every few minutes) to reach a diagnosis. A temporal-disruption manipulation presents the same information with natural hierarchical alignment (vitals nested within lab-result update cycles) versus misalignment (vitals and lab updates on competing, non-nested schedules).

**Subjects:** Medical residents at two levels of expertise—early-stage (Year 1–2) and advanced (Year 4–5)—plus a novice comparison group of pre-clinical students (total N ≈ 90, approximately 30 per group).

**Data Collection:** Diagnostic accuracy and time, self-reported cognitive load, concurrent eye-tracking (fixation patterns and transitions between fast and slow information sources), and think-aloud verbal protocols coded for temporal chunking strategies (e.g., grouping vital-sign trends into clinically meaningful episodes).

**Analysis:** A group × alignment interaction is predicted: novices should show large performance and load decrements under temporal misalignment, while advanced residents should show minimal effects. To test the chunking-mediation hypothesis, verbal protocols are coded for instances of temporal compression (re-describing fast events in terms of slower clinical patterns). Eye-tracking scan-path data are submitted to recurrence quantification analysis: experts under misalignment should show scan-path recurrence patterns resembling those of all participants under hierarchical alignment, evidence that experts internally reconstruct the temporal hierarchy that the task disrupted, effectively re-sculpting the attractor landscape through expertise-driven compression. Mediation analysis then tests whether the frequency of temporal chunking strategies statistically accounts for the expertise advantage, confirming that restoring hierarchical organization is the mechanism through which experts manage cross-timescale demands.